\documentclass[a4paper,12pt]{article}

\usepackage[latin2]{inputenc}
\usepackage{graphicx}
\usepackage{amstext}
\usepackage[]{ntheorem}
\usepackage{fancyhdr}
\usepackage{amsfonts}
\usepackage{amssymb}
\usepackage{amsmath,chemarrow}
\usepackage{array}
\usepackage{multirow}

\usepackage{t1enc}
\usepackage{tikz}
\usepackage{hyperref}
\usepackage[affil-it]{authblk}

\title{Quantitative supply security related significance measures for gas reservoires}

\author{D\'{a}vid Csercsik}

\affil{P\'{a}zm\'{a}ny P\'{e}ter Catholic University\\ Faculty of Information Technology and Bionics\\ Pr\'{a}ter~u. 50/A 1083 Budapest, Hungary \\
              Tel.: +36-1 886 47 00
              Fax: +36-1 886 47 24\\
              \emph{csercsik@itk.ppke.hu}}

\begin{document}

\maketitle

\abstract{Computational models corresponding to supply security in natural gas networks aim to describe flows and consumption values in the case of component failures or unforseen pipeline shutdowns. The role of natural gas reservoires in this process has only been marginally analyzed in such models, and typically only on the level of countries. In this paper we define a computational framework, which is capable of interpreting real flow and reservoir data to assign a quantitative supply security related measure to reservoires, depending on how much the given reservoir is critical in the process of restoring consumption outages in the network.}

\section{Introduction}
\label{Introduction}

Investments and developments related to natural gas infrastructure are considered as top priority for every country. The cost of such projects is significant,
typically comparable to the national GDP. The two main aspects taken into account during the planning of natural gas network or reservoir developments are (I) the expected economic benefits of the project, and (II) the potential effects regarding the security of supply.

The expected economic effect, the short and long term returns of gas-infrastructure developments can be estimated only in the context of the complex, regional financial and geopolitical environment (in other words, the international market for natural gas). At first glance it may be interesting, but economic benefits of a new pipeline are not necessarily related to the realized physical transport on the new transport route. The Velke Zlievce interconnector between Hungary and Slovakia for example \cite{mivsik2017eastring} allows Hungary to access western-European trading plattforms (HUBs). This potential access to alternative sources improved the bargaining positions of Hungary and resulted in the lower prices of Russian gas export to Hungary. The physical usage of the new interconnector is practically negligible ever since, but its economic effects are significant. Several non-cooperative \cite{holz2008strategic,egging2019global} and cooperative strategic models \cite{hubert2011investment,sziklai2020impact} aim to describe the economic aspects of infrastructure developments related to natural gas networks.

In contrast to economic analysis, which (typically) assumes normal operation of the infrastructure, the perspective of supply security focuses on scenarios, when the operation of the infrastructure and the network flows are negatively affected by external factors. The underlying causes may be of technical nature (as failures of pipeline elements or compressor stations), but they may be related to political disputes as well, as in the case of the 2006 and 2009 Ukrainian-Russian gas crises \cite{stern2006russian,stern2009russo}.
In the case of such events, when the consumption of certain network nodes is reduced due to disruptions in the transportation, we may distinguish two important elements in the restoration process. The first is the potential re-routing of gas available from sources already used during the normal operation of the network, and the other is the activation of additional sources as gas reservoires or previously unused LNG terminals in order to mitigate the damage.

Although the literature related to supply security of natural gas is quite extensive -- for publications relevant for the European region one may refer to \cite{stern2002security,correlje2006energy,holz2016role} --, studies using computational modelling tools and numerical simulations are less widespread \cite{scotti2012supply,villada2013simulation}. Moreover, even among quantitative models related to supply security, there are only a few studies analyzing the role of gas reservoires \cite{bavsova2016securing}, and these papers typically study the problem on the level of countries \cite{esnault2003need}, not on the level of the continental network.
The importance of a gas reservoir in a supply-disruption event depends on multiple factors. First, in addition to the total capacity of the reservoir, the maximal outlet rate limits the quantity of gas which may be withdrawn from a reservoir during a given time period (e.g. in a week or month). Second, the gas withdrawn from the reservoir must be transported to the nodes, where the consumption needs to be restored, thus free pipeline capacity must be accessible for the transport. As the available capacity depends on pipeline capacities and baseline flows as well, the availability (and thus the importance) of a given gas reservoir depends on the actual operational state of the network as well.

Our aim in this paper is to develop quantitative methods in order to model the re-routing and reservoir-activation scenarios taking place during the restoration process, and to define quantitative measures for the general supply-security related significance of reservoires of natural gas.

As the current paper focuses on the physical flows in supply security scenarios, no prices and costs are considered in this study: Neither the price of the production/transport of natural gas, nor the price of disruption of consumption is taken into account. Only physical flows and their limiting factors (pipeline and source capacities) are considered.

\section{Computational model}
\label{section_MM}
In this section we describe the concept of the proposed network-flow oriented framework. We consider the effects of transport disruption, and restoration processes on the time scale of one month. This time interval may be regarded as relevant for computational supply-security studies, since after one month, technical problems or political disputes causing disruptions in the network operations are usually relieved. In addition, this time detail also fits the data available on network flows, which are important inputs of the model in applications. The proposed computational principles may be however easily extended to longer or shorter periods.

\subsection{Basic concepts of the model}
\label{sec_basic_concepts}

The natural gas pipeline network is represented in the model as a directed acyclic graph with $n$ nodes and $m$ edges.
If we assume that no flow direction may be altered and no counter directed flows are allowed to take place, the acyclic property of the network graph may be regarded as plausible, considering the current modelling aim and context.

The nodes of the network are characterized by the following quantities:
\begin{itemize}
  \item Maximal monthly inlet value in million cubic meters (mcm): This value represents non-reservoir type sources of natural gas, as production sites (wells), LNG terminals etc., assuming normal opration. This value is denoted by $\bar{I}_j$ for node $j$.
  \item Monthly potential reservoir inlet value (in mcm): This value represents reservoirs, from which natural gas may be withdrawn. This value (denoted by $\bar{R}_j$) is equal to the volume of natural gas, which may be withdrawn from the reservoir in node $j$.
  \item Nominal monthly consumption value (in mcm): This value represents the monthly gas consumption of the node under normal circumstances (no supply disruption). We assume that this variable also means an upper bound for consumption. This value is denoted by $\bar{C}_j$ for node $j$.
\end{itemize}

Each edge $i$ of the network is characterized by a capacity value denoted by $\bar{f}_i$.

\subsubsection{Example network}

\begin{figure}[h!]
  \centering
  \includegraphics[width=10cm]{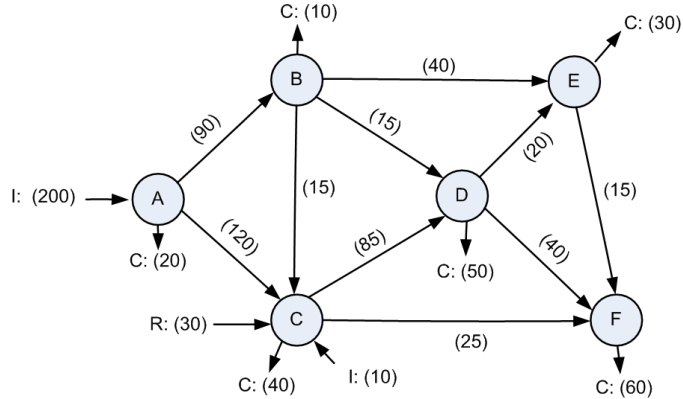}~
  \caption{Simple example network with 6 nodes. $I$ corresponds to inlets: the numbers in parentheses correspond to the maximal monthly inlet value of the respective node ($\bar{I}_j$). $R$ corresponds to reservoires: the numbers in parentheses correspond to the monthly potential reservoir inlet value of the respective node ($\bar{R}_j$). $C$ corresponds to consumption: the numbers in parentheses correspond to the nominal monthly consumption value of the respective node. Numbers in parentheses on edges correspond to maximal capacity ($\bar{f}$). Only nonzero values are included in the figure.}\label{Netw_example_1}
\end{figure}

As an example (motivated by the simple network described in \cite{scotti2012supply}), let us consider a 6-node ($n=6$) network, depicted in Fig
\ref{Netw_example_1}.

We denote the nodes of the network by A, B, C, D, E and F respectively.
We also summarize the nodal parameters of the network in Table \ref{node_params_example_N1}.

\begin{table}[h!]
\begin{center}
\begin{tabular}{|c|c|c|c|c|}
  \hline
  index & node notation&$\bar{I}_j$ & $\bar{R}_j$ & $\bar{C}_j$ \\ \hline
  1 & A & 200 & 0 & 20 \\
  2 & B & 0 & 0 & 10 \\
  3 & C & 10 & 30 & 40 \\
  4 & D & 0 & 0 & 50 \\
  5 & E & 0 & 0 & 30 \\
  6 & F & 0 & 0 & 60 \\
  \hline
\end{tabular}
\end{center}
\caption{Node parameters of the example network \label{node_params_example_N1}.}
\end{table}

The network has 10 edges ($m=10$), the parameters of which (capacity values) are summarized in Table \ref{edge_params_example_N1}.

\begin{table}[h!]
\begin{center}
\begin{tabular}{|c|c|c|c|}
  \hline
  index & from node & to node & \raisebox{-2pt}{$\bar{f}$} \\ \hline
  1 & A & B &90  \\
  2 & A & C &120 \\
  3 & B & C &15  \\
  4 & B & D &15  \\
  5 & B & E &40  \\
  6 & C & D &85  \\
  7 & C & F &25  \\
  8 & D & E &20  \\
  9 & D & F &40  \\
  10 & E & F & 15 \\
  \hline
\end{tabular}
\end{center}
\caption{Edge parameters of the example network \label{edge_params_example_N1}.}
\end{table}

\subsubsection{Operation modes of the network}

In the context of our modelling computations, we consider four different operation modes of a given network:

\begin{itemize}
  \item \emph{Normal operation mode} (NOM) describes the base case flows, inlets and consumptions in the network. We will assume that these values are defined prior.
  \item \emph{Disrupted operation mode} (DOM) of the network will describe, how a disruption of a line or a node will affect the inlets, consumption values and flows of the network.
  \item \emph{Re-routed operation mode} (RROM) of the network aims to describe the first reaction to the disruption: Gas available from various inlets is re-routed on available paths to replace the gas volumes in nodes where the disruption caused decrease in the consumption.
  \item \emph{Reserve-activated operation mode} (RAOM) describes the activation of reservoir sources, after the re-routing has taken place: Gas stored in reservoires is unloaded and routed on the available line capacities to further mitigate the effects of the disruption.
\end{itemize}

\paragraph*{State variables of the model:}
The state of the network in any given mode of operation is described with the following variables.

\begin{itemize}
\item Node-related variables: For each node $j$ the actual inlet value ($I_j \leq \bar{I}_j$), actual reservoir inlet value ($R_j \leq \bar{R}_j$) and the actual consumption value ($C_j \leq \bar{C}_j$) are considered as state-variables. We only consider supply-security usage of reservoires, thus we assume that reservoir inlets may be nonzero only in RAOM.
\item Edge-related variables: Gas flows are represented by a flow vector $f \in \mathcal{R}^m$. We assume that every flow is nonnegative, which means that the direction of flows must coincide with the direction of the edge. We assume that the unit of vector $f$ is million cubic meters (mcm). Furthermore $f \leq \bar{f}$.
\end{itemize}

For every operation mode, we require that inflow and outflow must be in balance for each node. After a motivational example, which demonstrates these network states and highlights the possible measures applicable for the significance of reservoires in this model context, we will describe in detail how the DOM and the following RROM and RAOM may be calculated consecutively, based on the network parameters and on the prior given state variables corresponding to the NOM.

\subsection{A demonstrative example of the disruption-restoration process}
\label{demo_example_1}

\begin{figure}[h!]
  \centering
  \includegraphics[width=8.2cm]{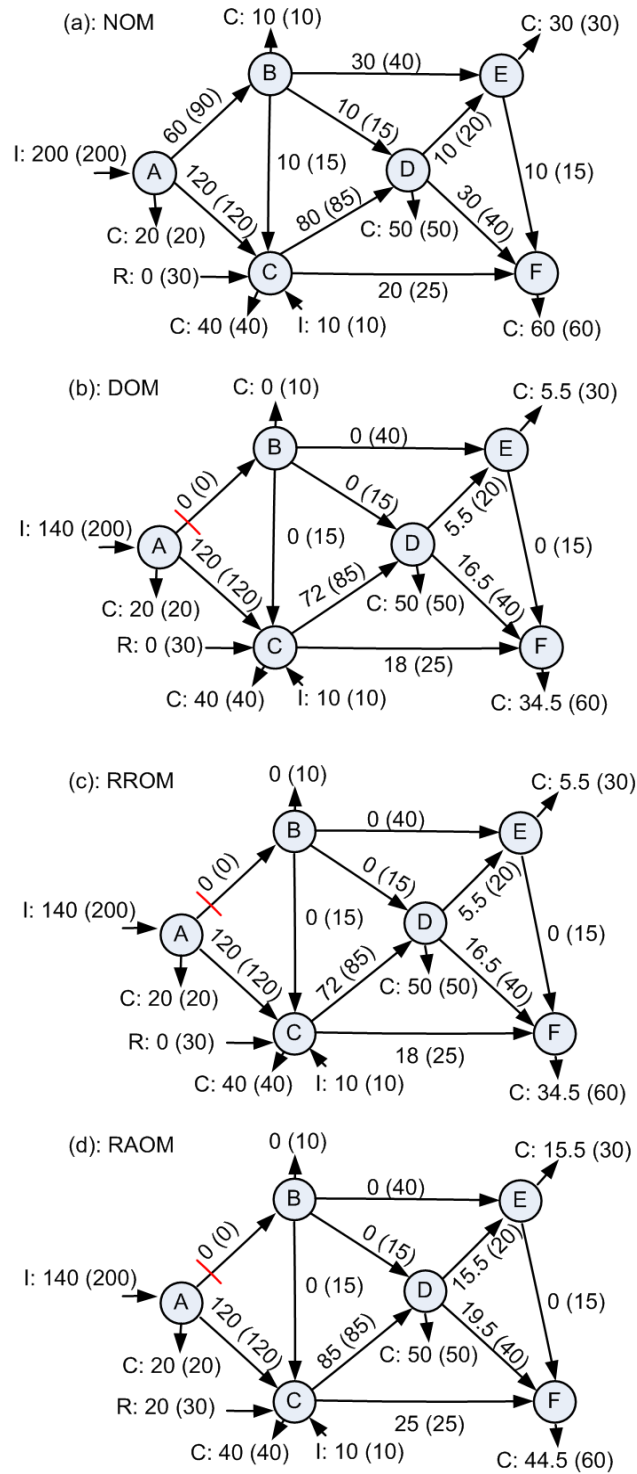}~
  \caption{4 states of the example network. For each inlet ($I$), reservoir inlet ($R$) and consumption value ($C$), the actually realized values are indicated without parentheses, while the numbers in parentheses correspond to the maximal/ideal values (as before). The notation is the same for line flows: numbers without parentheses denote the realized values while numbers in parentheses are the maximal values (i.e. capacities).}\label{example_NDRR_1}
\end{figure}

In this subsection we demonstrate how the 4 network states are realized, using the simple network depicted in Fig \ref{Netw_example_1}.
Figure \ref{example_NDRR_1} depicts the 4 states of the network.

Fig. \ref{example_NDRR_1} (a) depicts the NOM of the network. As already mentioned, in the context of the paper we assume that the state variables corresponding to this state (nominal inlet, consumption and flow values) are given prior. We can see that in NOM, all consumption needs are fulfilled in the network, and nodal balances hold for each node.

Fig. \ref{example_NDRR_1} (b) depicts the DOM of the network. In this case, we assume the disruption of line 1 (between node A and node B).
 The inlet of A is reduced according to the amount originally transported on line 1 (60 units).
 Since no gas arrives in node B, it is straightforward that its outflows and consumption are zeroed out. In the case of nodes, which are also affected by the disruption, but still have incoming gas flows (like C, for which the input is decreased by 10 units), following \cite{scotti2012supply}, we assume that
\begin{itemize}
  \item Each node prioritizes its own consumption: From the inflows the nodal consumption is covered first, and the rest is forwarded.
  \item The ratio of the forwarded gas volumes should be equal to the respective ratios in NOM.
\end{itemize}
Regarding node C, these assumptions imply that in the DOM, node C first covers its own consumption needs (40 units), and forwards the remaining gas to nodes D and F, according to the constraint regarding the proportions: $80/20=72/18$.
The same applies for node D: First, the own consumption is covered and the remaining gas volumes are forwarded to nodes E and F according to the original proportions ($10/30=5.5/16.5$).

Fig. \ref{example_NDRR_1} (c) depicts the RROM of the network. In this state, available (non-reservoir type) inlets are activated and the corresponding volumes are routed on free line capacities to compensate for the outages in the nodes affected by the disruption (in this case E and F). Let us note that during the model calculations we will assume that flows in the DOM must be unaffected by the re-routing -- in other words, only available inlet volumes may be rerouted on available capacities (which are determined by the flows remaining in the DOM).

In this particular case however, only node A has available inlets, but the outgoing capacities of A are fully exploited, so this potential additional volume is not able to reach the nodes affected by the disruption (B, E and F). In other words, in the case of this example, no practical re-routing takes place,

Fig. \ref{example_NDRR_1} (d) depicts the RAOM of the network. In this case, the reservoir in node C is activated, and additional 20 units
of gas is routed to E and F to compensate for the outage. The outward capacities of node C limit the activation of the reservoir (more gas can not be routed to E and F). We assume that the restoration volumes from the reservoires are routed to achieve the most egalitarian compensation possible. In this case, perfect balance is possible (in the terms of node E and F, since B is not reachable from C): node E and node F both receive 10 units of gas from the reservoir in node C.

Our aim in this paper is to assign significance measures to gas reservoires which reflect how much their network position and capacity allows
them to compensate for disruption-implied demand reductions. In this case, we can say that the total consumption outage implied by the considered disruption (after re-routing -- which in his particular case implied no explicit flown modifications ) was 60 units (10 units in B, 24.5 units in E and 25.5 units in F), from which the activation of the reservoir in node C was able to compensate 20 units. In this case, the reservoir of C compensated for 33.3 \% of the consumption reduction.

Of course the above calculation corresponds only to the failure of the line A-B. If we average the above value for the failures of each line (if line failure probabilities are available we can also take the respective weighted average), the result may be interpreted as a simple significance measure for the reservoir in question.

\subsubsection{Effect of network expansion}

\begin{figure}[h!]
  \centering
  \includegraphics[width=10cm]{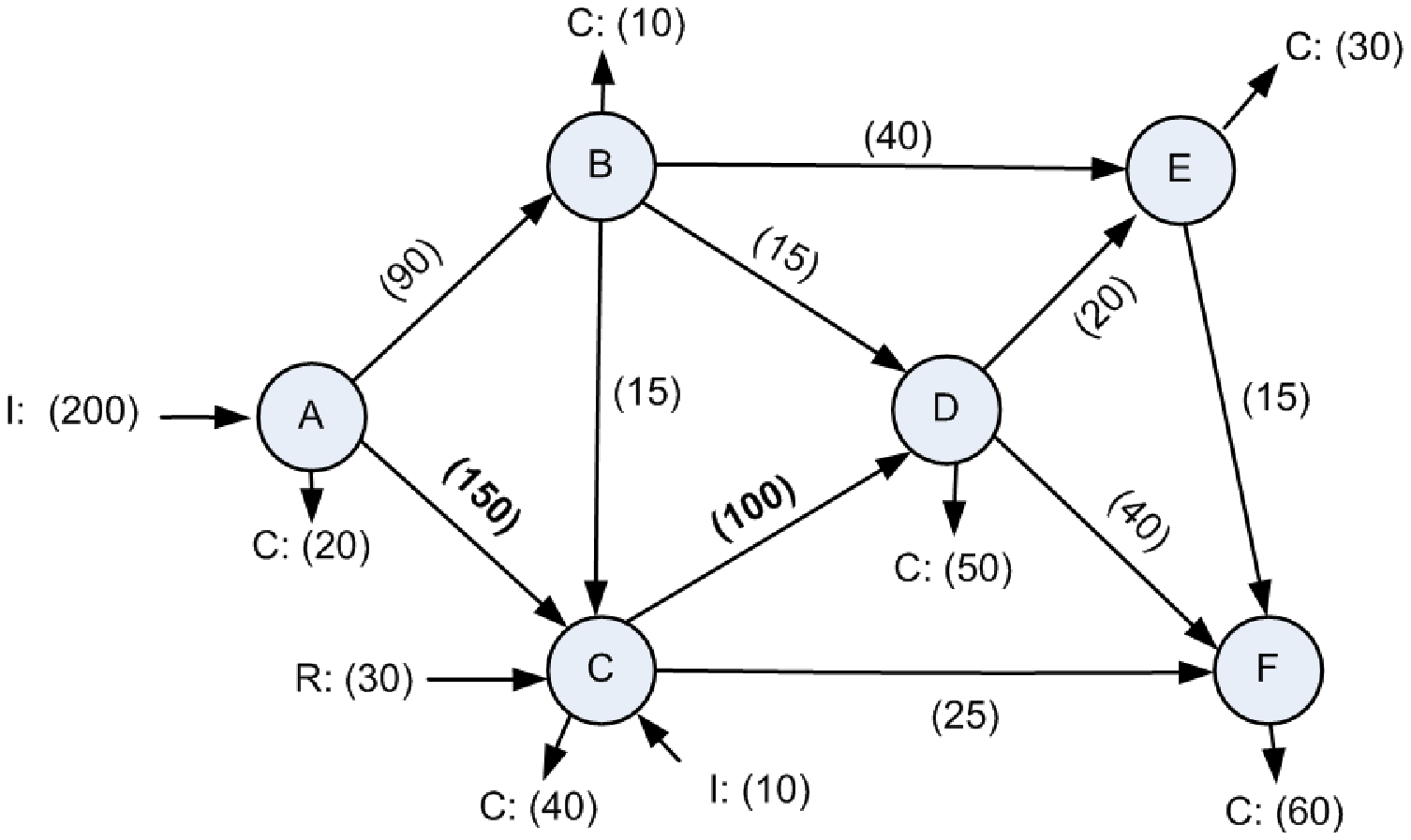}~
  \caption{The simple example network with extended capacities. Modified values are emphasized with bold typeface.}\label{Netw_example_2}
\end{figure}

In this subsection we use the previously introduced simple example to show how network expansion can modify the modelled process, and the resulting measures. Let us assume that the nodal parameters of the network remain unchanged, but the capacity of the edges 2 (from A to C) and 6 (from C to D)
are increased from 120 to 150 and 85 to 100 respectively. The updated edge parameters are summarized in Table \ref{edge_params_example_N2},
and the modified network is depicted in Fig \ref{Netw_example_2}.

\begin{table}[h!]
\begin{center}
\begin{tabular}{|c|c|c|c|}
  \hline
  index & from node & to node & \raisebox{-2pt}{$\bar{f}$} \\ \hline
  1 & A & B &90  \\
  2 & A & C &\textbf{150} \\
  3 & B & C &15  \\
  4 & B & D &15  \\
  5 & B & E &40  \\
  6 & C & D &\textbf{100}  \\
  7 & C & F &25  \\
  8 & D & E &20  \\
  9 & D & F &40  \\
  10 & E & F & 15 \\
  \hline
\end{tabular}
\end{center}
\caption{Edge parameters of the example network with extended capacities. Modified values are emphasized with bold typeface.  \label{edge_params_example_N2}}
\end{table}

\begin{figure}[h!]
  \centering
  \includegraphics[width=8cm]{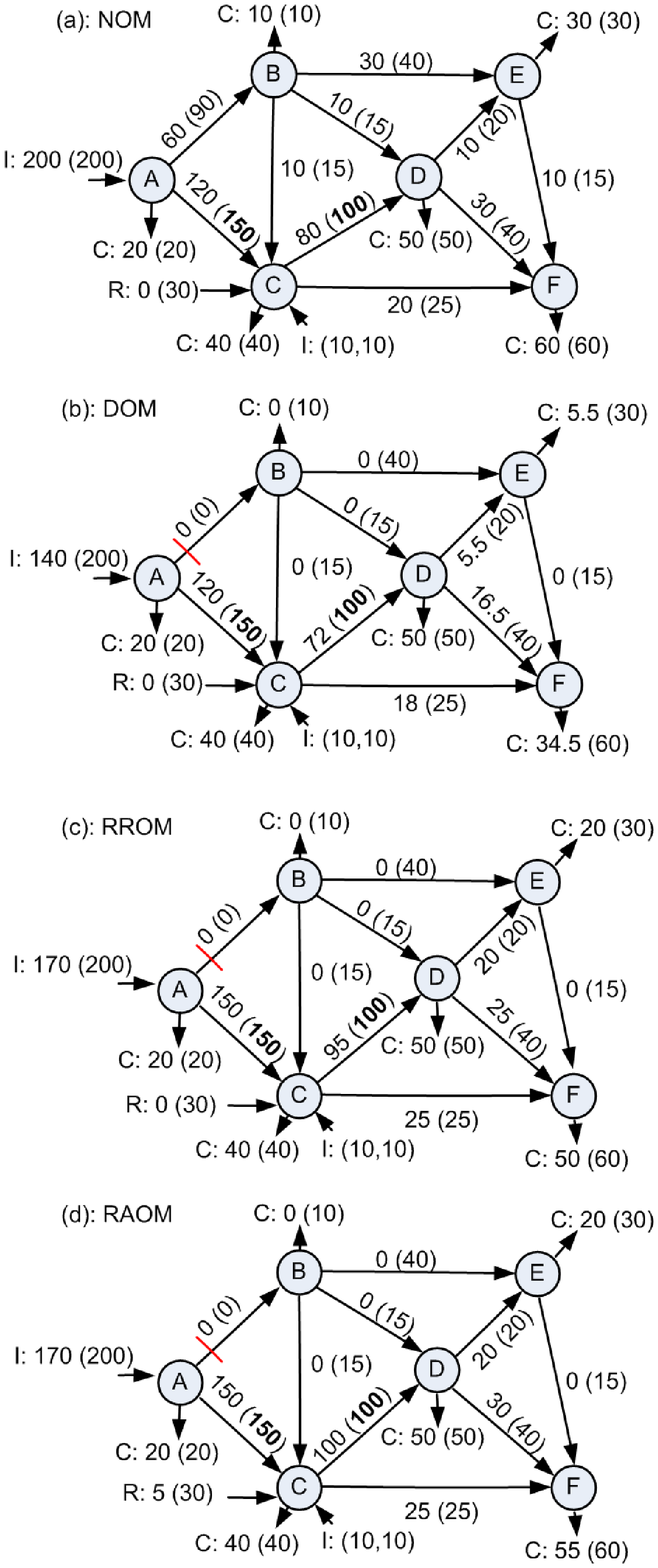}~
  \caption{4 states of the extended example network. The parameters, which have been increased due to the extension are denoted with bold typeface. For each inlet ($I$), reservoir inlet ($R$) and consumption value ($C$), the actually realized values are indicated without parentheses, while the numbers in parentheses correspond to the maximal/ideal values (as before). The notation is the same for line flows: numbers without parentheses denote the realized values while numbers in parentheses are the maximal values (i.e. capacities).}\label{example_NDRR_2}
\end{figure}

We assume furthermore, that the NOM of the network is the same as before (depicted in Fig. \ref{example_NDRR_1} (a)).
We consider again the same disruption, regarding line 1 (from A to B).
The process, and the resulting 4 network states are depicted in Fig. \ref{example_NDRR_2}.

In Fig. \ref{example_NDRR_2} (a) and (b) we can see that the base-case flows and the effect of the disruption are the same as before in
Fig \ref{example_NDRR_1}.

In Fig. \ref{example_NDRR_2} (c) however, in contrast to the original case depicted in Fig \ref{example_NDRR_1}, we can see that re-routing has real significance in his case: In the RROM, an additional 30 units of gas from node A is re-routed to nodes E and F, using the available capacities of the network, which are present due to the capacity extensions. Although in subsection \ref{details_of_computation} we will see the detailed computational formulation of the re-routing process, the principles of the re-routing calculations are as follows.

We consider the nodes, where the DOM resulted in consumption outage, and which are reachable from the nodes with available additional inlet.
We consider the outage values, which we aim to compensate, and look for a re-routing of available gas, which results in maximal overall compensation value (the sum of the remaining outages over the network must be minimal). As this solution, however, may not be unique, in the second step, from the possible outage-minimizing solutions regarding accessible outage nodes, we choose the one, which is the closest to 'equal compensation' in the absolute sense. As in general, this consideration does also not necessary imply a unique solution, in the third step we determine the setup which uses the shortest available routes.

In this particular case, 30 units are re-routed from node A. Node B is not accessible, thus we are interested only in nodes E and F. Node E has an outage of 24.5 units, while node F has an outage of 25.5 units. The equal compensation in the absolute sense would be if 15 - 15 units of the 30 units would be transferred to E and F. In this case, this is however not possible, so we choose the most close feasible solution: We transfer 14.5 to E via D and 15.5 to F directly.

Fig. \ref{example_NDRR_2} (d) depicts the activation of reserves. In this case, the free capacities allow 5 units of gas to be transferred from the reservoir of node C to node F.

In this case, we can say that the total consumption outage implied by the considered disruption (after re-routing -- which in his particular case implied no explicit flown modifications ) was 30 units (10 units each in B, E and F), from which the activation of the reservoir in node C was able to compensate 5 units. The reservoir of C compensated for 16.667 \% of the consumption reduction in this case.

As we can see, the modification of the network affected (more precisely reduced) the significance of the reservoir in the context of its consumption-reduction potential. In the following we describe the details of the calculations corresponding to the determination of the network states corresponding to the various operation modes.

\subsection{Details of the computational formulation}
\label{details_of_computation}

As we mentioned before, we assume that the NOM of the network is given prior.
In the following we describe how the network states in further operation modes (DOM, RROM and RAOM) are derived from this initial state.

\subsubsection{DOM calculations}
\label{DOM_calculations}
In the framework of the proposed model we consider line failures. Any line of the network may fail, and, according to our consumptions, this means
that the maximal transport capacity of the line in question is reduced. Basically, we assume that the capacity of the line is reduced to 0, but partial failures may be also considered -- the proposed computational methods may be applied in this case as well.
This reduction in the transfer capacity typically also induces the reduction of line flows, as in the case of Fig. \ref{example_NDRR_1} (b), where the flow of the line A-B has been reduced to 0.

This initial reduction of a given flow induces imbalances in some of the network nodes (i.e. the start node and end node of the edge in question). In order to determine the modified state variables (inlets, consumptions and line flows) of the network which resolve this imbalance and constitute the DOM state, we perform the following steps for each imbalanced node.

\begin{itemize}
  \item If the node has surplus (the sum of inlets and inflows exceeds the sum of consumption and outflows), inlet and inflow values are decreased in a way which conserves the proportions of inlet and inflow values as much as possible.
  \item If the node has deficiency (sum of consumption and outflows exceeds the sum of inlets and inflows), consumption and outflow values are decreased. In this case however we assume that each node with deficiency gives priority to own consumption: Such nodes aim to cover own consumption first and the outflows are updated according to the remaining gas quantity.
\end{itemize}

After performing these steps, two possibilities may arise.

\begin{itemize}
  \item There are no more imbalanced nodes in the network. In this case the calculations are finished, and the resulting states constitute the DOM.
  \item One or more imbalanced nodes are still present in the network. In this case, the calculations are repeated.
\end{itemize}

\begin{figure}[h!]
  \centering
  \includegraphics[width=7cm]{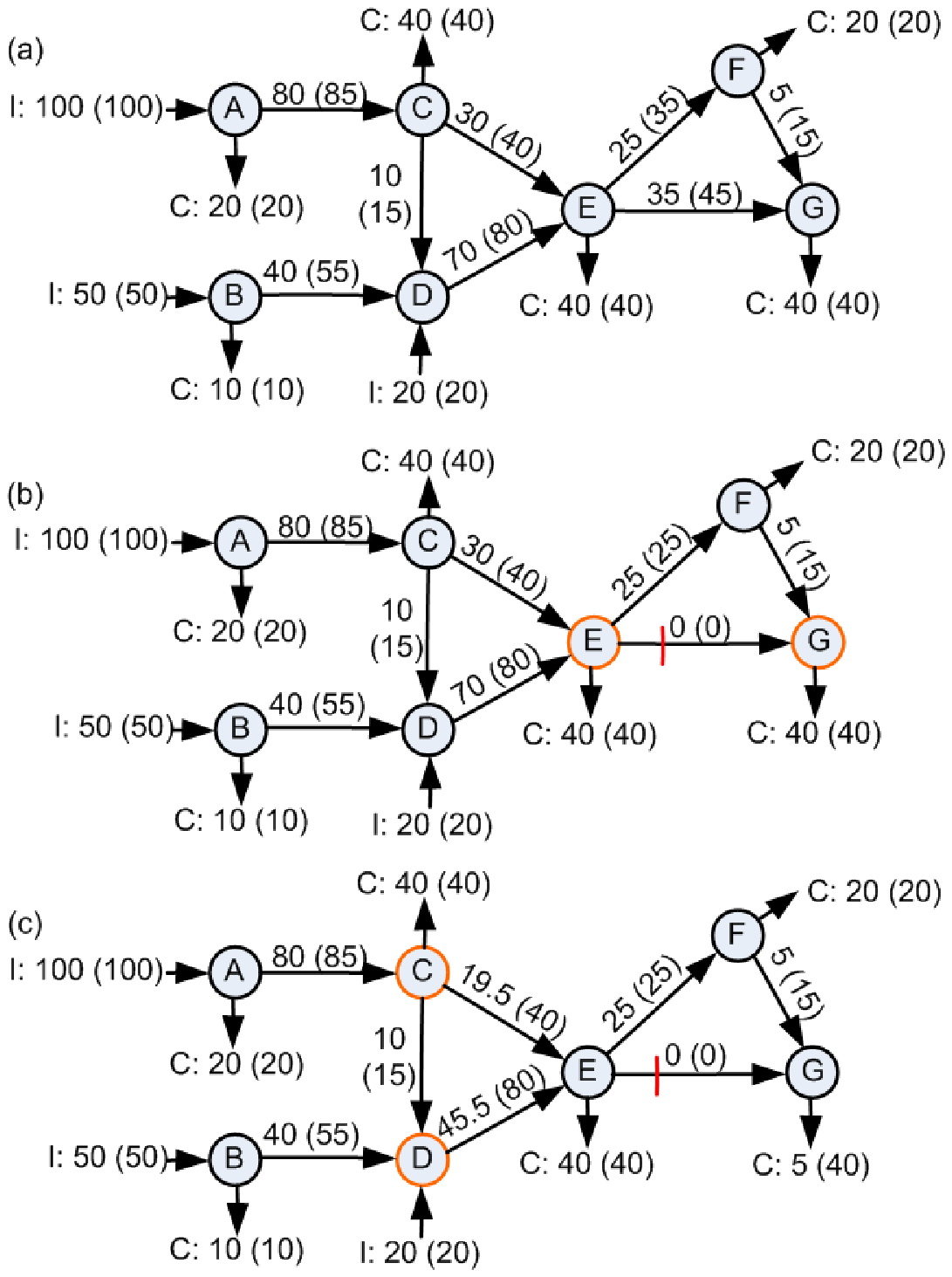}~
  \includegraphics[width=7cm]{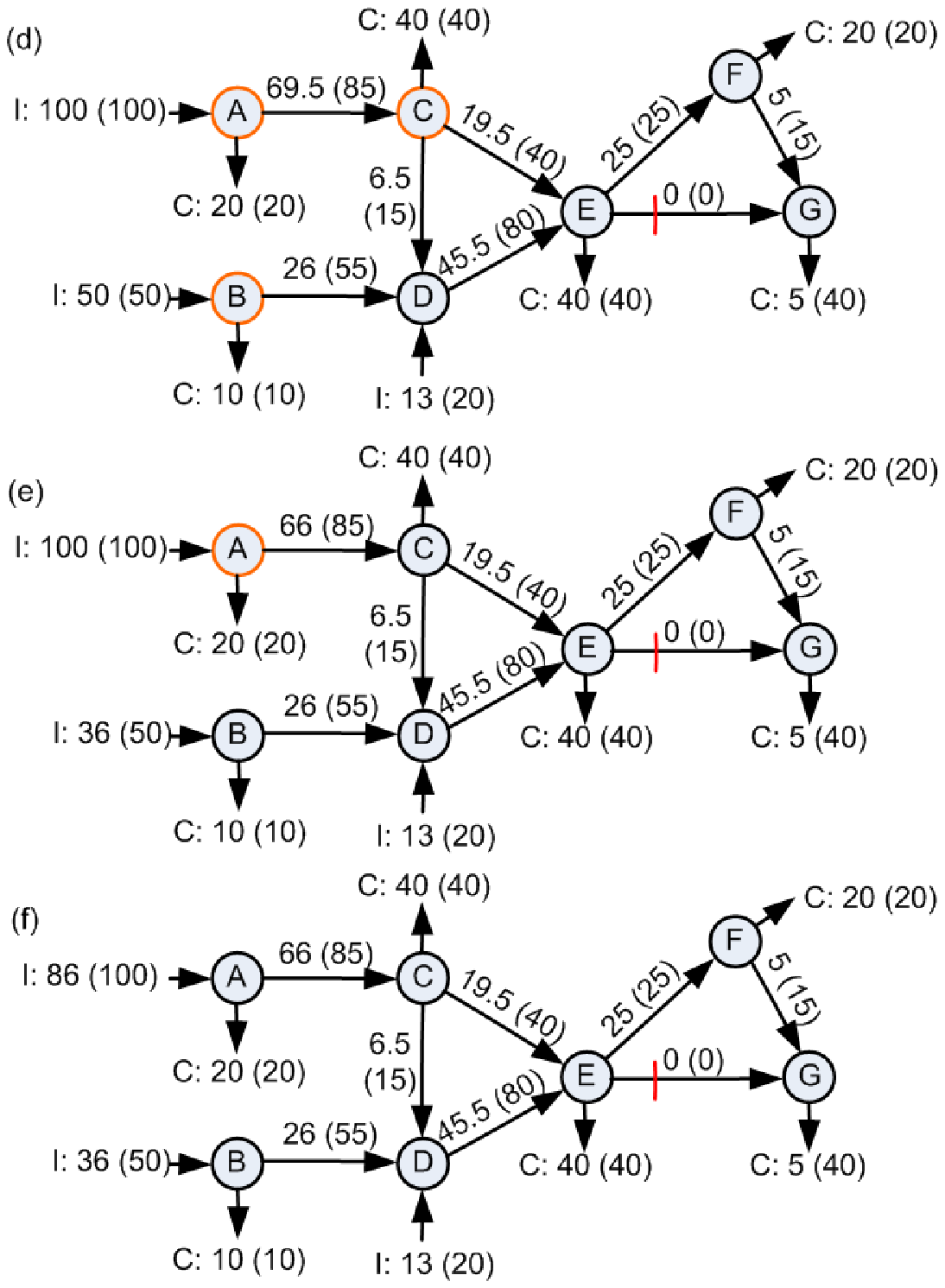}
  \caption{Iterative calculation of the DOM state in an example network. Imbalanced nodes are denoted with orange.}\label{Fig_DOM}
\end{figure}

The above iterative process is demonstrated on a less trivial example in detail (step-by-step) in Fig. \ref{Fig_DOM}, where nodes with imbalance are highlighted. In this figure we can see that in the first step (b) the initial line failure affects the starting node and the end node of the failed line (namely E and G). In the next step (c), the balance of these nodes is restored according to the principles described above, but this implies further imbalance in nodes C and D. As the process progresses, in the end (f), all nodes are in balance again, and the network sates corresponding to the DOM are determined. The acyclicity of the graph guarantees that the process will stop after a finite number of iterations.

\subsubsection{RROM and RAOM calculations}

The calculation of the states corresponding to RROM and RAOM is performed similarly, the only difference is that
while in the case of RROM the unused capacity of (normal) inlets are considered as sources, in the case of RAOM the reservoir capacities are considered as sources. In the following we describe the RROM calculations -- the RAOM can be calculated accordingly, mutatis mutandis.

If we compare the actual consumption values of the DOM with nominal consumption values we can determine the set of nodes (nodes with consumption outage), which are affected by the disruption (in Fig. \ref{example_NDRR_1} these are nodes C E and F). Following this we determine that subset of these outage nodes, which are reachable from any of the source nodes: An outage node (B) is reachable from the source node (A), if a directed path from A to B exists, along which all the lines have nonzero free capacity (capacity over the actual flow). We denote the set of reachable nodes by
$N^R$.

The calculation of the states corresponding to the RROM can be described as a sequence of flow-optimization processes. The variables of these flow-optimization problems are the same in each step (inlets of the sources, line flows and consumption values), but the constraints and the objective function is different in each step. The steps considered are as follows.

\paragraph*{Maximizing the total restoration of consumption:} In this step the following optimization problem is considered. For each affected node $j$ (nodes for which the consumption has been decreased as a result of the line fault), the reduction in consumption can be defined as $C^-_j=C_j^{NOM}-C_j^{DOM}$ where $C_j^{NOM}$ is the consumption of node $j$ in NOM and $C_j^{DOM}$ is the consumption of node $j$ in DOM (the first is given prior, and the second is determined via the method described in subsection \ref{DOM_calculations}).
The available additional inlets considered in the re-routing process are defined as $\bar{I}_j^{RR}=\bar{I}_j-I_j^{DOM}$ for each node $j$, where $I_j^{DOM}$ denotes the inlet of node $j$ in the DOM.
The available line capacities for each line $i$ in the re-routing process are defined as $\bar{f}_i^{RR}=\bar{f}_i-f_i^{DOM}$, where $f_i^{DOM}$ denotes the flow of line $i$ in the DOM.

The variable vector of the optimization corresponding to the first step of re-routing is defined as in Eq. (\ref{state_vector_RR_1}).

\begin{equation}\label{state_vector_RR_1}
  x^{RR}=\left(
      \begin{array}{c}
        I^{RR} \\
        C^{RR} \\
        f^{RR} \\
      \end{array}
    \right)
\end{equation}

$I^{RR} \in \mathcal{R}^n$ stands for the additional inlets realized (activated) in the re-routing process,
$C^{RR} \in \mathcal{R}^n$ denotes the vector of additional consumption values, resulting from the re-routing. These additional consumptions aim to mitigate the consumption-reductions resulting from the disruption.
$f^{RR} \in \mathcal{R}^m$ denotes the flows corresponding to the re-routing process.

The constraints of the problem are described in equation (\ref{optim_con_RR_1}), while eq. (\ref{obj_fncn_optim_RR}) formulates the objective function: The aim in this step is to restore as much consumption as possible. We can see that the optimization of this step results in a linear programming problem.

\begin{align}
I^{RR}_j ~\leq~ \bar{I}_j^{RR}~~~~~~\forall~j \nonumber \\
C^{RR}_j ~\leq~ C^-_j~~~~~~\forall~j \nonumber \\
f^{RR}_i ~\leq~ \bar{f}_i^{RR}~~~~~~\forall~i \label{optim_con_RR_1}
\end{align}

\begin{equation}\label{obj_fncn_optim_RR}
\max_{x^{RR}}~~~ \sum_j C^{RR}_j
\end{equation}

Let us denote the obtained maximal value of the objective function by $C^{RR}_{T1}$.

\paragraph*{Equalizing the consumption-restoration over reachable nodes:} As the set of $x^{RR}$ vectors maximizing the total restoration of consumption may be unique, in the next step we formalize the consideration, that we prefer such vectors, which equally reduce the consumption reduction in each reachable node. This consideration is formulated as a quadratic programming problem, in which the variable vector is the same as before (see Eq. (\ref{state_vector_RR_1})), the constraints (\ref{optim_con_RR_1}) still hold, but we add the constraint (\ref{optim_con_RR_2}) as well, and modify the objective function as described in Eq. (\ref{obj_fncn_optim_RR_2}) where the indices i1, i2, i3,... correspond to nodes in $N^R$.

\begin{equation}\label{optim_con_RR_2}
\sum_j C^{RR}_j =C^{RR}_{T1}
\end{equation}

\begin{equation}\label{obj_fncn_optim_RR_2}
\min_{x^{RR}}~~~ (C^{RR}_{i1}-C^{RR}_{i2})^2 + (C^{RR}_{i2}-C^{RR}_{i3})^2 + ...
\end{equation}

Let us denote the $C^{RR}$ vector resulting from the optimization problem by $C^{RR}_1$.

\paragraph*{Minimal usage of network lines:} It is possible that the solution obtained for $x^{RR}$ in the previous step is still not unique.
In this case we add the constraint (\ref{optim_con_RR_3}) to the problem, and minimize the linear objective function (\ref{obj_fncn_optim_RR_3}) to choose the flows which use the minimal number of edges in the network.

\begin{equation}\label{optim_con_RR_3}
C^{RR} =C^{RR}_{T1}
\end{equation}

\begin{equation}\label{obj_fncn_optim_RR_3}
\min_{x^{RR}}~~~ f^RR \cdot 1
\end{equation}

The above consideration does not distinguish between the transportation costs among edges, but
if transfer costs for the edges are available, this step may be modified accordingly.

\section{Discussion}
\label{sec_Discussion}
As it has been described in subsection \ref{sec_basic_concepts}, the proposed method assumes an acyclic flow pattern as input (corresponding to the NOM). This assumption of acyclicity deserves some discussion.
It seems straightforward to assume that because of the transfer costs, cyclic flows are contra-productive and unnecessary in the network, so one may think that this assumption naturally holds for real data.
In contrast, considering the realistic European pipeline network, the flows observable in any instance are results of complex financial transactions composed of agreements for various terms, which are sometimes binding (e.g. it is possible that according to a long-term agreement gas must be transferred from east to west on a pipeline, even if on the actual gas is cheaper on western hubs). This may result in cyclic flows on the network.
On the other hand, data available about pipeline gas flows is cumulated (e.g. for months). It is possible that daily acyclic flows result in monthly cumulated flows, which do have cycles.

If the proposed computational framework is used to interpret real flow data, this issue may be solved by a preprocessing step of, which removes the
cyclic flows from the input.

\section{Conclusions and future work}
\label{sec_Conclusions}

In this paper, we provided a computational framework, which assigns supply security related significance measures to gas reservoires.
The method uses the concept of nominal, disrupted, re-routed and reserve-activated operation modes (NOM, DOM, RROM and RAOM respectively), which are calculated iteratively. The assumptions used during the calculations try to avoid the prioritization of any consumer or line, and they are formulated according to the principle that both the effect of failures and restoration efforts take place without any prioritization among consumers or network lines.

During the DOM calculations described in subsection \ref{DOM_calculations}, the effect of flow disruptions are iteratively back-propagated through the network, assuming proportional decrease of nodal inflows (we require the inflow proportions to be constant in every affected node).
During the forth-propagation of flow disruptions, it is assumed that every node prioritizes its own consumption and sends the remaining gas further, aiming to keep the proportion of nodal outflows -- this assumption is in accordance with \cite{scotti2012supply}.

During the RROM and RAOM calculations, the first principle is always to restore as much consumption as possible, and the second aim is to do this via the most equal support of affected nodes, which are reachable from additional sources.

According to the above calculations in the case of a line failure, we can determine how much of the consumption reduction may be restored based on the emergency activation of the reservoir in question. Averaging these values for every possible line failure considered results in a quntitative supply security related significance measure for the corresponding gas reservoir.

\subsection{Future work}

As discussed earlier, the acyclicity assumption regarding the NOM flows is a critical element of modelling assumptions. In the future, in addition to the iterative method described in subsection \ref{DOM_calculations}, it would be desirable to develop additional approaches, which may be used for the determination of DOM in a network, even in the case of cyclic flows present. An optimization-based approach could possibly substitute the iterative algorithm. The details of this alternative DOM calculation will be described in the future.

\section{Acknowledgements}

This work has been supported by the \'{U}NKP-20-5 New National Excellence Program of the Ministry for Innovation and Technology from the source of the National Research, Development and Innovation Fund.

\newpage

 \bibliographystyle{plain}
  \bibliography{SS}

\end{document}